\documentclass[aps,prl,amsmath,amssymb,twocolumn,superscriptaddress]{revtex4-1}
\usepackage{graphicx}
\usepackage{dcolumn}
\usepackage{bm}
\usepackage[utf8]{inputenc}
\usepackage[english]{babel}
\usepackage{amsfonts}
\usepackage{hyperref}
\usepackage{physics}
\usepackage{epstopdf}
\usepackage{mathtools}
\usepackage[capitalise]{cleveref}
\usepackage{float}
\usepackage[space]{grffile}
\usepackage{color}
\usepackage[normalem]{ulem}
\usepackage{subfigure}

\binoppenalty=\maxdimen
\relpenalty=\maxdimen

\addto\captionsenglish{}

\begin{document}

\title{Enhancing cavity QED via anti-squeezing:  synthetic ultra-strong coupling}

\author{C. Leroux}
\affiliation{Department of Physics, McGill University, Montr\'eal, Qu\'ebec, Canada.}
\author{L. C. G. Govia}
\affiliation{Institute for Molecular Engineering, University of Chicago, Chicago, Illinois, USA.}
\author{A. A. Clerk}
\affiliation{Institute for Molecular Engineering, University of Chicago, Chicago, Illinois, USA.}

\begin{abstract}
We present and analyze a method where parametric (two-photon) driving of a cavity is used to exponentially enhance the light-matter coupling in a generic cavity QED setup, with time-dependent control.  Our method allows one to enhance weak-coupling systems, such that they enter the strong coupling regime (where the coupling exceeds dissipative rates) and even the ultra-strong coupling regime (where the coupling is comparable to the cavity frequency).  As an example, we show how the scheme allows one to use a weak-coupling system to adiabatically prepare the highly entangled ground state of the ultra-strong coupling system.  The resulting state could be used for remote entanglement applications.
\end{abstract}

\maketitle

%%%%%%%%%%%%%%%%%%%%%%%%%%%%%%%%%%%%%%%%%%%%%%
\emph{Introduction-- }Cavity QED (CQED), the interaction between a single two-level system (qubit) and a quantized mode of a cavity, is a ubiquitous platform \cite{Walther:2006aa} that has widespread utility, ranging from the study of fundamental physics \cite{Haroche2006}, to the cutting edge of quantum information \cite{McKay:2015aa,OMalley:2016aa,Wang2016}.  The most interesting regimes of CQED correspond to a qubit-cavity coupling that is strong enough to dominate dissipation rates.  While this has been achieved in several architectures, e.g.~\cite{Brune:1987aa,Thompson:1992aa,Wallraff:2004aa,Khitrova2006,Viennot:2015aa,Mi:2017aa,Stockklauser2017}, in many others \cite{Schuster:2010aa,Kubo:2010fk,Petersson:2012aa,BienfaitA:2016aa} it remains extremely challenging. Even more challenging is reaching the so called ultra-strong coupling (USC) regime, where the coupling strength is comparable to the qubit/cavity frequency. Here, counter-rotating terms cannot be ignored, and the system is best described by the quantum Rabi model \cite{Rabi:1936aa,Rabi:1937aa}, which is known to exhibit a wide range of interesting phenomena, such as strongly entangled and nonclassical eigenstates \cite{Solano2012,2010Nori,Braak:2016aa,Gheeraert:2017aa,Leroux2017}. To date, only specially designed architectures have reached USC experimentally \cite{Beltram2009,Huber2009,Gross2010,Mooij2010,Yoshihara:2017aa}, though simulations of USC have also been considered \cite{Solano2012,Mezzacapo2014,DiCarlo2016}.

In this paper, we show how simple detuned parametric driving of a cavity can be used to dramatically enhance the effective qubit-cavity coupling in a generic CQED system.  This can be used to turn a weakly coupled system into a strongly coupled one, and even push one from strong coupling to the USC regime.   After describing the general idea, we show it enables the direct study of Rabi-model physics in a system whose bare coupling is far from the USC regime.  Further, the ability to turn on and off the coupling enhancement leads to new applications.  We show how our approach allows one to leverage the entanglement of the USC ground state to generate remote entanglement between the qubit and a traveling waveguide mode.  This is a key ingredient in quantum teleportation \cite{Bennet93} and teleported gates \cite{Gottesman1999}, quantum repeaters \cite{Gisin2011}, and quantum networks \cite{Kimble2008}. Note that related approaches for coupling enhancement have been considered in the context of cavity optomechanics \cite{Nori2015,Clerk2016}, a system with a markedly different kind of light-matter interaction.

\begin{figure}
    \centering
    \includegraphics[width=0.18\textwidth]{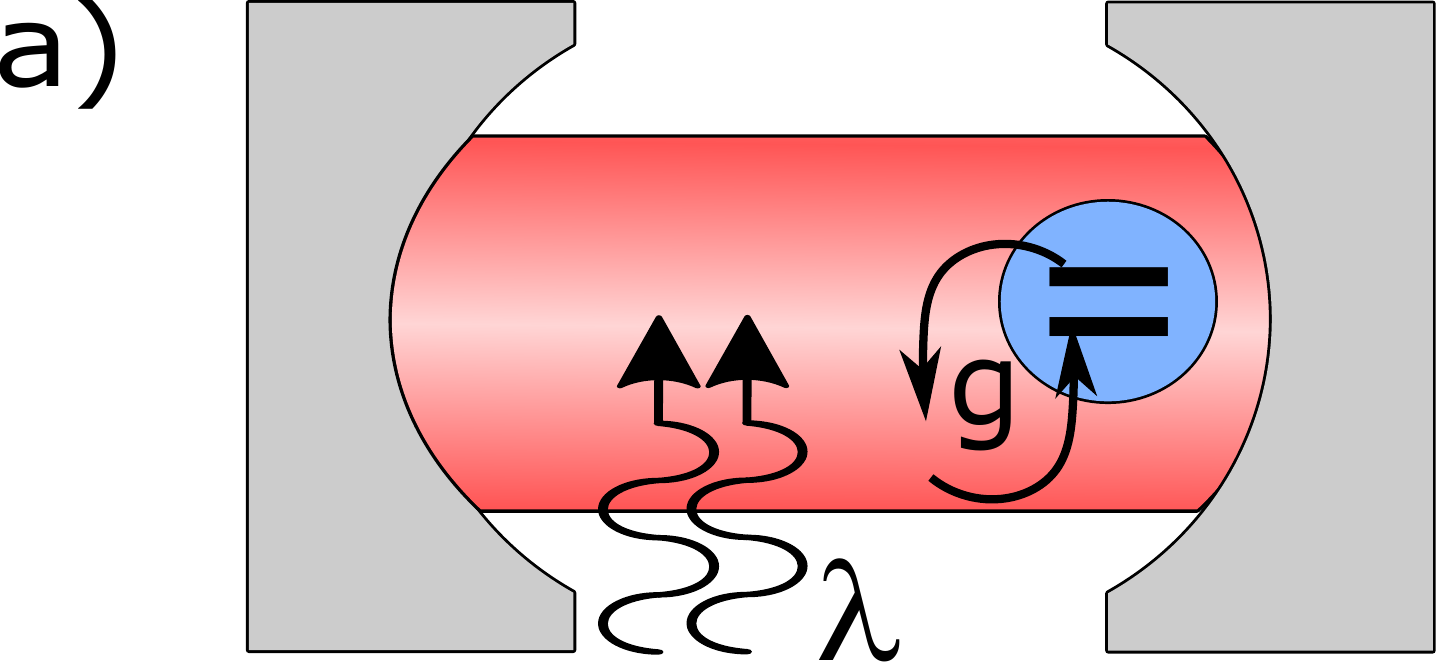}
    \space \space
    \includegraphics[width=0.24\textwidth]{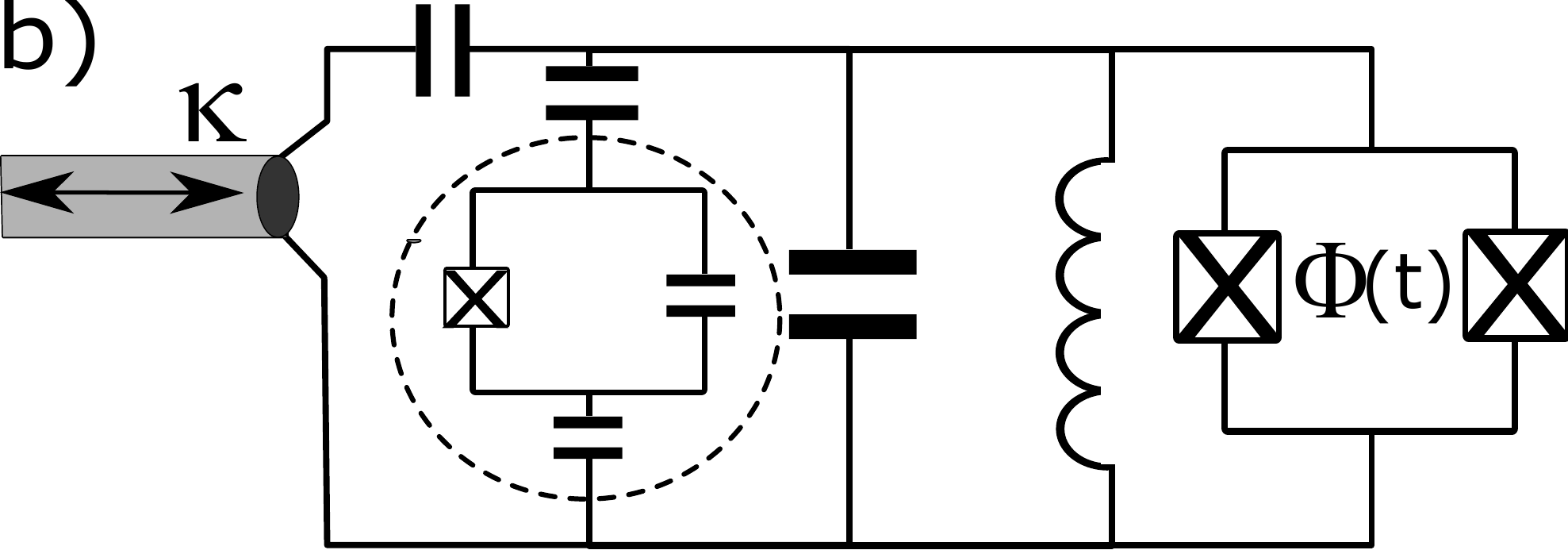}
    \caption{a) Schematic of the setup: a parametrically driven cavity (drive strength $\lambda$ and drive frequency $\omega_p$) is weakly coupled to a qubit with rate $g$. b) A potential realization in circuit QED: a qubit coupled capacitively to a lumped-element cavity. A flux-pumped SQUID connected to the cavity implements the parametric drive, and the cavity couples to a transmission line with rate $\kappa$.}
    \label{fig:RemoteEntanglement}
\end{figure}

%%%%%%%%%%%%%%%%%%%%%%%%%%%%%%%%%%%%%%%%%%%%%%
\emph{Model-- }We start by considering a qubit weakly coupled to a cavity, with the cavity subject to a two-photon (i.e.~parametric) drive, see \cref{fig:RemoteEntanglement}.  Working in a frame rotating at half the parametric drive frequency $\omega_p/2$, the Hamiltonian is
\begin{align}
	\hat{H}(t) = \delta_c \hat{a}^{\dagger}\hat{a} + \frac{\delta_q}{2}\hat{\sigma}_z - \frac{\lambda(t)}{2} ( \hat{a}^{\dagger 2}+\hat{a}^2) +
	g(\hat{a}^{\dagger}\hat{\sigma}_-+\hat{\sigma}_+\hat{a}),
\label{eq:HrotSim}
\end{align}
where $\hat{a}$ is the cavity annihilation operator and $\hat{\sigma}_{\pm}$ are qubit raising/lowering operators. $\lambda(t)$ is the time-dependent parametric drive amplitude, $g$ is the qubit-cavity coupling strength, and $\delta_{c/q} = \omega_{c/q} - \omega_p/2$, are the cavity and qubit detunings (with $\omega_{c/q}$ being the cavity/qubit frequencies).  The weak value of $g$ implies that the qubit-cavity interaction is well-described by the excitation-conserving Jaynes-Cummings coupling written above.  Note that a parametric drive can be implemented in many different physical architectures.  For example, in circuit QED, one can modulate the flux through a SQUID embedded in the cavity (see e.g.~\cite{Yamamoto:2008aa}). In what follows, we will be interested exclusively in detuned parametric drives with $|\delta_c| > \lambda$, which ensures that \cref{eq:HrotSim} is stable.

%%%%%%%%%%%
% FIG 1 Schematic
%%%%%%%%%%%
%%%%%%%%%%%

The instantaneous cavity-only part of $\hat{H}(t)$ can be diagonalized by the unitary $\hat{U}_{\rm S}[r(t)] = \exp\left[r(t)\left(\hat{a}^2-\hat{a}^{\dagger 2}\right)/2\right]$, where the squeeze parameter $r(t)$ is defined via
\begin{equation}
	\tanh 2r(t) = \lambda(t)/\delta_c.
	\label{eq:rDefn}
\end{equation}
The Hamiltonian in the time-dependent squeezed frame described by $\hat{U}_{\rm S}[r(t)]$ is
\begin{align}
	&\nonumber \hat{H}^{\rm S}(t) \equiv
		 \hat{U}_{\rm S}[r(t)]\hat{H}\hat{U}_{\rm S}^\dagger[r(t)] - i \hat{U}_{\rm S}\dot{\hat{U}}_{\rm S}^{\dagger}\\ &
	\phantom{\hat{H}^{\rm S}(t)} = \hat{H}_{\rm Rabi}(t) + \hat{H}_{\rm Err}(t) + \hat{H}_{\rm DA}(t), \label{eq:HSt}
\end{align}
where
\begin{align}
&\nonumber \hat{H}_{\rm Rabi}(t) = \Omega_c[r(t)] \hat{a}^{\dagger}\hat{a} + \frac{\delta_q}{2}\hat{\sigma}_z + \frac{g}{2}e^{r(t)} (\hat{a}^{\dagger}+\hat{a})(\hat{\sigma}_++\hat{\sigma}_-),\\
&\nonumber \hat{H}_{\rm Err}(t) = - \frac{g}{2}e^{-r(t)} (\hat{a}^{\dagger}-\hat{a})(\hat{\sigma}_+-\hat{\sigma}_-), \\
& \hat{H}_{\rm DA}(t) = - \frac{i\dot{r}(t)}{2}\left(\hat{a}^{\dagger 2}-\hat{a}^2\right)
	\label{eqs:Hams}.
\end{align}
The Hamiltonian $\hat{H}_{\rm Rabi}(t)$ has the form of the usual Rabi Hamiltonian, with an enhanced coupling $\tilde{g} = g e^{r(t)}/2$ and an effective cavity frequency $\Omega_c[r(t)] = \delta_c\sech 2r(t)$ (which decreases with increasing $r$).
As $e^{r(t)}$ becomes arbitrarily large as we approach the instability threshold $\lambda = |\delta_c|$, the effective qubit-cavity coupling in $\hat{H}_{\rm Rabi}$ can be orders of magnitude larger than the original coupling $g$.

The remaining terms in Eqs.~(\ref{eqs:Hams}) describe undesired corrections to the ideal Rabi Hamiltonian.  $\hat{H}_{\rm Err}(t) $ is explicitly suppressed by $e^{-r(t)}/2$, and thus is negligible in the large amplification limit $e^{r(t)} \rightarrow \infty$ (as long as the cavity population in the squeezed frame remains finite). % $\ll $e^{r(t)}$).
Note that the cavity vacuum in the squeezed frame corresponds in the original lab frame to a squeezed vacuum state with squeeze parameter $r(t)$.
The last correction term $\hat{H}_{\rm DA}(t)$ vanishes explicitly for a time-independent drive amplitude (and only plays a limited role in the adiabatic preparation scheme discussed later).

Thus, for large parametric drives, we have that our system is {\it unitarily equivalent} to the Rabi-model Hamiltonian $\hat{H}_{\rm Rabi}(t)$ with an exponentially enhanced coupling strength. This enhancement is a consequence of the coherent parametric drive modifying the eigenstates of the cavity Hamiltonian: these are now squeezed photons, whose amplified fluctuations directly yield a larger interaction with the qubit.  We stress that this enhancement is not equivalent to simply injecting squeezed light into the cavity (as this does not change the Hamiltonian).  It is also distinct from the usual $\sqrt{n}$ enhancement associated with the Jaynes-Cumming interaction between a qubit and an $n$-photon Fock state, as in the squeezed frame, the interaction is enhanced for both small and large photon numbers.

\emph{Weak to Strong Coupling-- }  The simplest application of our approach is to enhance the coupling in a weak-coupling CQED system (where $g$ is much smaller than the cavity damping rate $\kappa$).  Even if the resulting enhanced coupling $\tilde{g} < \kappa$, the increase could lead to dramatic enhancement of measurement sensitivity for spin or qubit detection, as the signal to noise ratio scales quadratically with $g$.   This could be of particular utility in systems involving electronic or nuclear spins coupled to microwave cavities, where couplings are naturally weak \cite{Schuster:2010aa,BienfaitA:2016aa,Mi:2017aa,Stockklauser2017}.  The enhancement of a dispersive qubit measurement in this regime (where $\tilde{g} < \kappa$) can in some cases equivalently be understood from the perspective of amplification, see Ref.~\onlinecite{Levitan:2016aa} for a full discussion.

%%%%%%%%%%%
% Fig 2 Vac Rabi Splitting
%%%%%%%%%%%
\begin{figure}[t]
    \centering
    \includegraphics[width=0.4\textwidth]{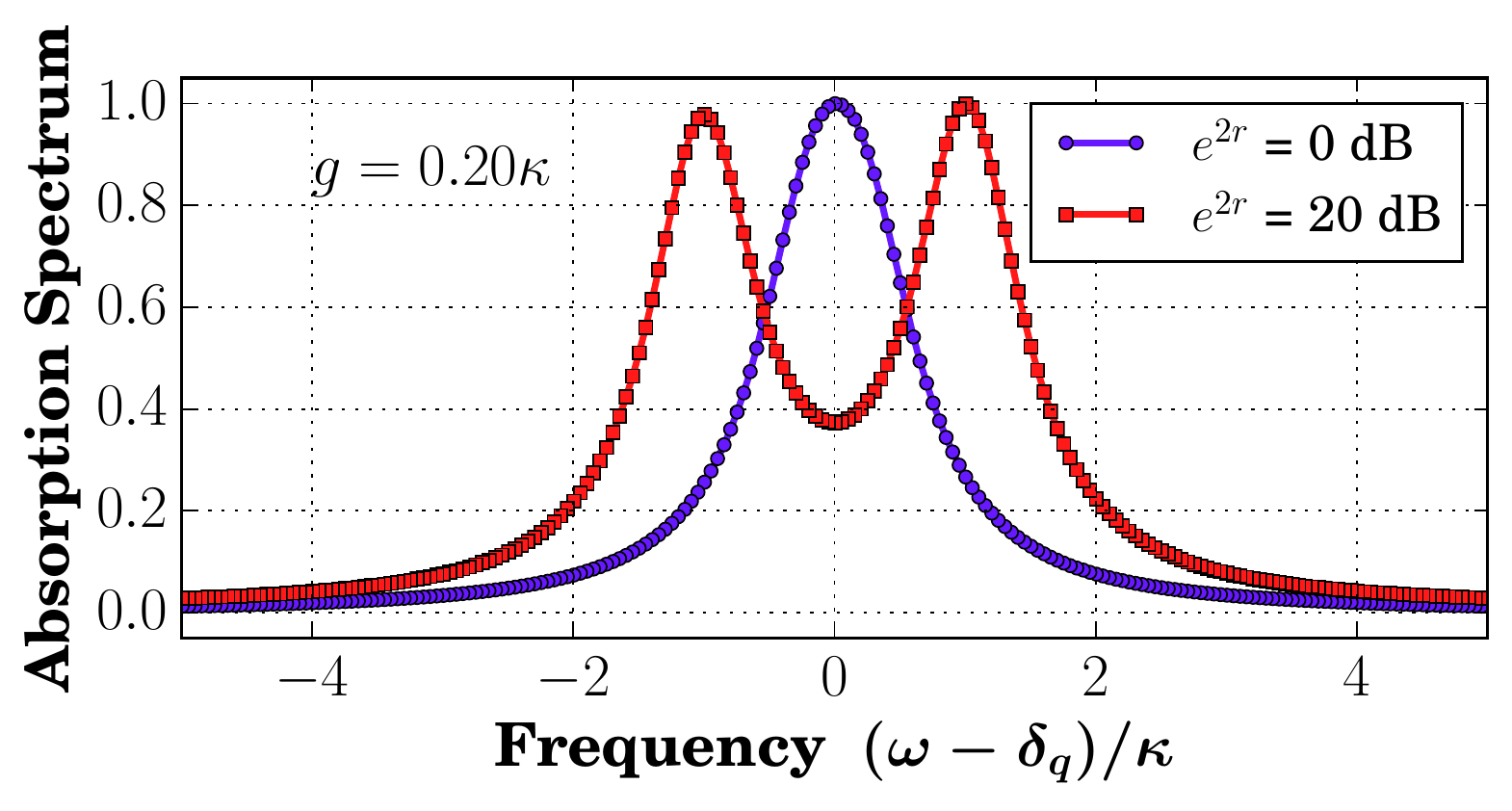}
    \caption{Qubit absorption spectrum as determined by a master equation simulation, for a resonant CQED system with $g = 0.2 \kappa$,
    $\kappa = \gamma = 5 \times 10^{-4}\delta_c$.  The blue (red) curve is for zero parametric drive (parametric drive corresponding to $e^{2r} = 20$ dB).  The qubit frequency in the lab frame $\delta_q$ is adjusted in both cases to maintain resonance with the cavity (i.e. $\delta_q = \Omega_c[r]$, c.f.~Eq.~(\ref{eq:HSt})).  For the red curve, we assume the cavity is driven by squeezed vacuum, such that the system is in vacuum in the squeezed frame. }
    \label{fig:VRS}
\end{figure}
%%%%%%%%%%%

Perhaps more interesting is the ability of our scheme to push a system with a weak bare coupling ($g < \kappa$) into the regime of strong effective coupling, $\tilde{g} \gtrsim \kappa$.  In this regime, we expect that the parametrically driven system will exhibit features of a true strong coupling CQED system.  A hallmark of strong coupling is vacuum Rabi splitting (VRS) \cite{Haroche2006}, where, e.g., the qubit absorption spectrum splits as a function of frequency (due to qubit-cavity hybridization).  In \cref{fig:VRS} we show the qubit absorption spectrum (obtained from a master equation simulation \cite{SupMat}) as a function of frequency for a bare coupling $g = 0.20 \kappa$, both with and without a parametric drive.  As expected, the coupling enhancement due to the drive leads to a clear VRS. Note that to obtain a simple zero-temperature spectrum, we assume that the cavity is driven by squeezed vacuum noise in the lab frame, which corresponds to simple vacuum noise in the squeezed frame used in Eq.~(\ref{eq:HSt}).  This ensures that the system starts in the ground state in the squeezed frame \cite{SupMat}.

Strong coupling in CQED enables a number of applications, ranging from nonlinear quantum optics at the two-photon level to single-atom lasing \cite{Khitrova2006}.  Parametric driving makes these accessible even in systems with a weak bare coupling.

\begin{figure}
    \centering
    \includegraphics[width=0.4\textwidth]{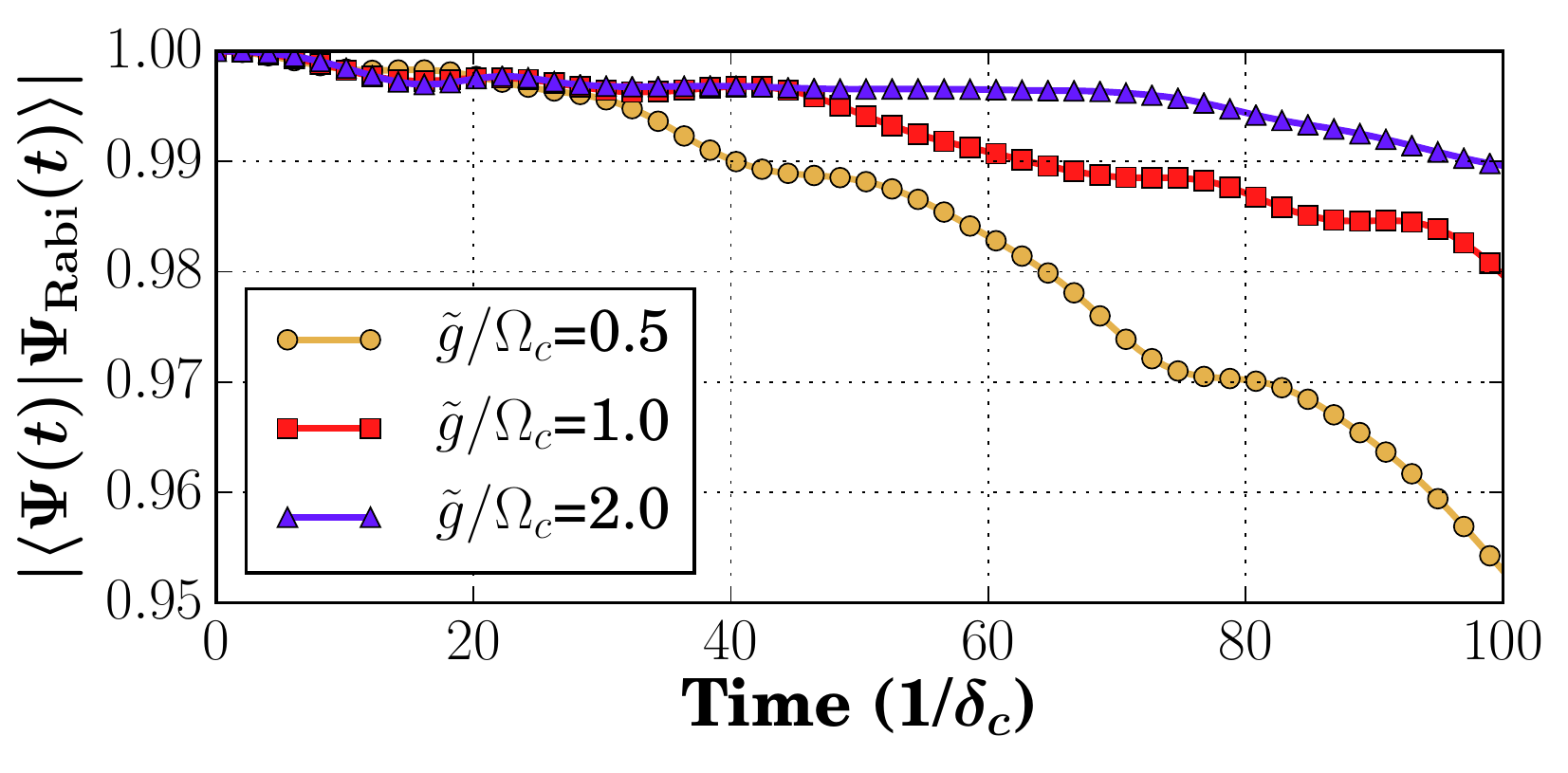}
    \caption{{\bf Simulation of a coupling quench.}  Time-dependent overlap between the state
    $\ket{\psi(t)}$ obtained in our synthetic USC simulation, versus the state $\ket{\psi_{\rm Rabi}(t)}$ obtained in ideal Rabi-model evolution; the system starts in both cases in the weak-coupling ground state.
   Different curves are for different  values of dimensionless enhanced coupling $\tilde{g} / \Omega_c \equiv g e^r/2\Omega_c$. $\delta_q = 0.1\delta_c$ and $g=0.05\delta_c$. Dissipation is neglected and the initial state is the vacuum $\ket{0,-z}$ in the squeezed frame.}
    \label{fig:dynam}
\end{figure}

%%%%%%%%%%%%%%%%%%%%%%%%%%%%%%%%%%%%%%%%%%%%%%
\emph{Dynamical Simulation of USC Quench-- }Parametric driving can even be pushed further, allowing a weak or strong coupling CQED system to be enhanced into the USC regime.  In Fig.~(\ref{fig:dynam}) , we show that our approach allows a faithful realization of the USC regime by comparing the dynamical evolution of the parametrically driven system (including all terms in  \cref{eq:HSt}) against a simulation of just the ideal Rabi Hamiltonian $\hat{H}_{\rm Rabi}$.  We start the system in the $g=0$ ground state of  \cref{eq:HSt}, and thus are simulating a quench-type protocol where the (ultra-strong) coupling is suddenly turned on.  Fig.~(\ref{fig:dynam}) plots the time-dependent fidelity between the simulated state and the ideal Rabi-model state, for several values of parametric drive strength.  The parametrically driven system faithfully reproduces the ideal Rabi-model evolution over long timescales.  The fidelity is even better for larger coupling enhancements, as the larger the squeezing, the more the suppression of the unwanted terms in $\hat{H}_{\rm Err}$ (c.f.~Eqs.~(\ref{eqs:Hams})).
%A weakly or strongly coupled system can also be enhanced to the USC regime, as we now show by comparing a dynamical simulation of the true Rabi Hamiltonian, $\hat{H}_{\rm Rabi}$, to the full \cref{eq:HSt} for constant $\lambda(t) = \lambda$. This includes the exponentially suppressed effects of the error term, but the diabatic term is zero for constant $\lambda$. As an initial state in the squeezed frame, we start with cavity vacuum and the qubit in its ground state. In the lab frame, this would correspond to the cavity in a squeezed state with squeezing parameter $-r$. We compare the state fidelity of the synthetic Rabi evolution of \cref{eq:HSt} with the true Rabi evolution for several values of the enhanced coupling strength, and this is shown in \cref{fig:dynam}. As can be seen, our synthetic simulation has high fidelity over a long evolution, especially for stronger enhancements since $\hat{H}_{\rm Err}$ (c.f. \cref{eq:HSt}) is exponentially suppressed.

%%%%%%%%%%%%%%%%%
%  Fig 3:  Dynamical simulation
%%%%%%%%%%%%%%%%%
%%%%%%%%%%%%%%%%%

%%%%%%%%%%%%%%%%%%%%%%%%%%%%%%%%%%%%%%%%%
\emph{Adiabatic preparation of entangled USC ground states-- }
In contrast to the above quench protocol, we can start with the trivial ground state of a weakly coupled CQED system (i.e.~ground state of \cref{eq:HrotSim} for $\lambda(t)=0$), and then adiabatically prepare the ground state of the ultra-strong coupling Rabi model by slowly ramping up the parametric drive amplitude $\lambda(t)$.  Further, once the desired state is achieved, the parametric drive can be turned off, returning the system to weak-coupling dynamics.

The ability to prepare strong coupling ground states and then return to weak coupling allows a number of useful protocols.  After preparation,
one could turn off the coupling and allow the cavity state to leak into the waveguide or transmission line coupled to the cavity, implying that any cavity-qubit entanglement is now qubit-propagating photon entanglement; this enables remote entanglement protocols.  Alternatively, as the protocol ends with the system in a weak coupling regime, the cavity state can be directly probed using standard weak-coupling techniques, for example by using the qubit  \cite{Bertet:2002aa,Hofheinz:2009aa,Vlastakis2013,Wang2016,Flurin:aa}. This addresses the long-standing issue of how to observe the non-trivial aspects of the ground state of the quantum Rabi model.

While one could use this approach to prepare the ground state of the Rabi Hamiltonian $\hat{H}_{\rm Rabi}(t)$ in any parameter regime, we focus on the case $\delta_q = 0, \tilde{g} \gtrsim \Omega_c$, where the ground state has the form of an entangled cat state:
\begin{equation}
	\ket{\Psi_{\rm Target}(t)} = \frac{\ket{\alpha(t)} \ket{+x} - \ket{-\alpha(t)}\ket{-x}}{\sqrt{2}}. \label{eq:ECS}
\end{equation}
Here $\ket{\alpha}$ denotes a coherent state in the cavity, and $\ket{\pm x}$ denote $\hat{\sigma}_x$ qubit eigenstates; the displacement $\alpha \propto \tilde{g}$ (see EPAPS for full expression) \cite{SupMat}.  Note that as this is the ground state in the squeezed frame, in the original lab frame the state will correspond to the qubit being entangled with squeezed, displaced cavity pointer states.  As discussed above, preparing this state and then turning off the parametric drive allows one to create a nontrivial entangled state where the qubit is entangled with a propagating (squeezed, displaced) wavepacket.  Unlike more standard approaches  (e.g.~\cite{Vlastakis:2015aa}), this is accomplished without any controls or drives applied directly to the qubit.

To consider the robustness of our approach, we simulate adiabatic state preparation in the presence of both cavity and qubit dissipation.  These are treated via a standard Linblad master equation, which in the original lab frame takes the form:
\begin{align}
	\dot{\hat{\rho}} = i \comm{\hat{\rho}}{\hat{H}(t)}+ \gamma\mathcal{D}[\hat{\sigma}_-] \hat{\rho} + \kappa \mathcal{D}[\hat{a}] \hat{\rho}, \label{eq:MER}
\end{align}
where $\mathcal{D}[\hat{x}] \hat{\rho} = \hat{x}\hat{\rho}\hat{x}^{\dagger}-\frac{1}{2}\acomm{\hat{x}^{\dagger}\hat{x}}{\hat{\rho}}$, $\kappa$ is the cavity damping rate, $\gamma$ is the intrinsic qubit decay rate, and we have assumed zero temperature environments. Note that the simple form of this master equation is justified by the fact that we have a driven system with a large drive frequency $\omega_p$ (see EPAPS \cite{SupMat}); as a result, complications associated with strong-coupling master equations \cite{Beaudoin11} do not apply.

We parameterize the time-dependent parametric drive amplitude $\lambda(t)$ via $r(t) = r_{\rm max} \tanh\left(t/2\tau\right)$ and Eq.~(\ref{eq:rDefn}), where $\tau$ sets the effective protocol speed, and $r_{\rm max}$  is the final maximum value of the time-dependent squeeze parameter.  The evolution runs from $t=0$ to $t = t_f \gg \tau$.  We quantify the success of the protocol using the fidelity $F(t)$ between the dynamically generated state and the final desired target state in \cref{eq:ECS},
\begin{equation}
F(t) = \sqrt{\matrixel{\Psi_{\rm Target}(t_f)}{\hat{\rho}^{\rm S}(t)}{\Psi_{\rm Target}(t_f)}},
\label{eq:F(t)}
\end{equation}
where $\hat{\rho}^{\rm S}(t)$ denotes the system density matrix in the squeezed frame. Achieving a good fidelity involves picking a value of $\tau$ that is large enough to ensure adiabaticity, but not so large that dissipative effects corrupt the evolution.

%%%%%%%%%%%%%%%
% Fig. 4:  Adiabatic state prep
%%%%%%%%%%%%%%%%
\begin{figure}[t]
    \centering
    \includegraphics[width=0.4\textwidth]{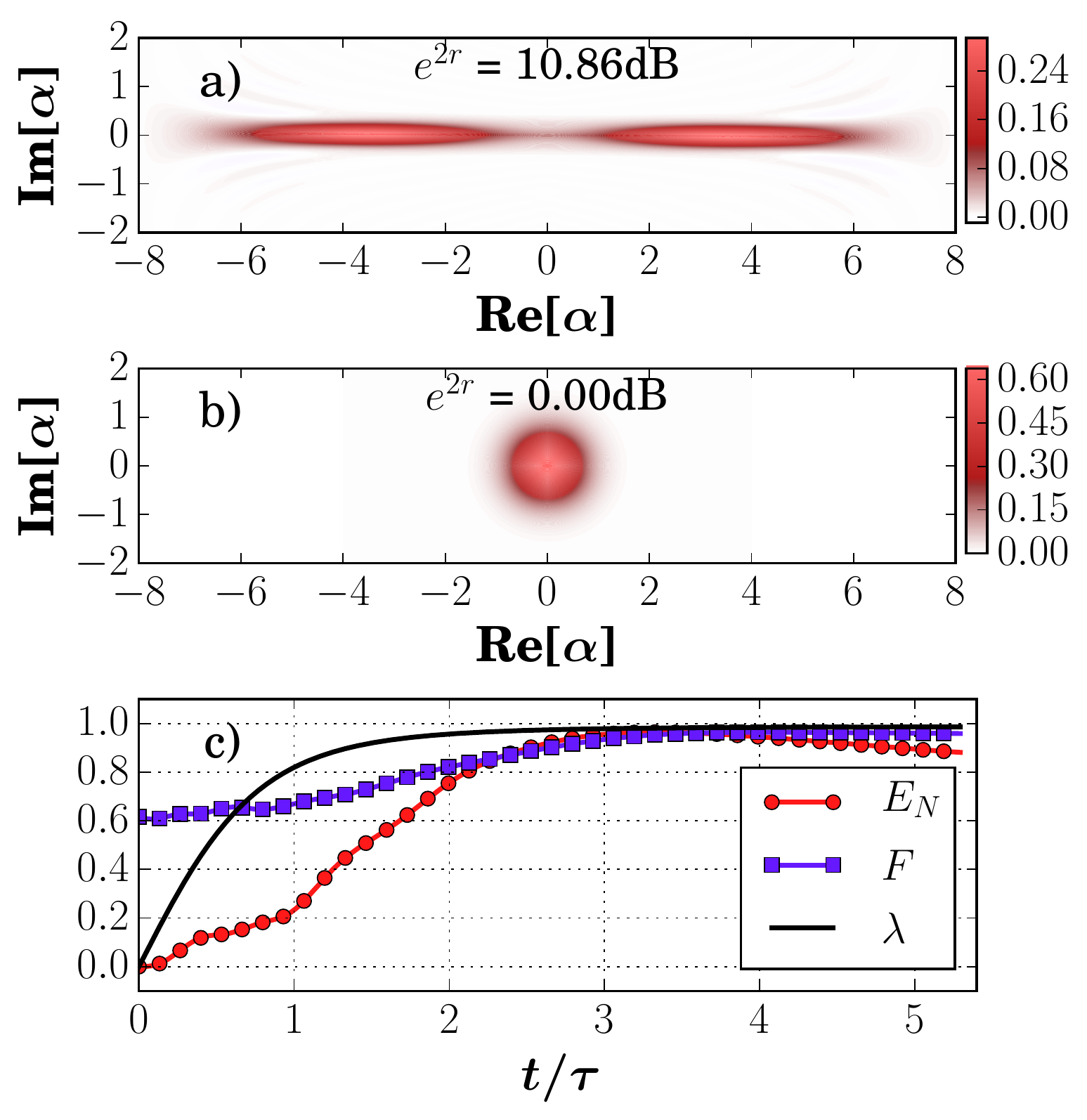}
    \caption{{\bf Adiabatic preparation of entangled ultra-strong coupling ground state.}
    (a)  A CQED system with $g = 0.1 \delta_c$, $\delta_q = 0$ is initially prepared in the weak coupling ground state
    $\ket{\Psi^{\rm S}(0)}=\ket{0}\ket{-z}$.  The parametric drive is turned on adiabatically (see main text) at a rate $1/\tau = 0.01 \delta_c$, to a final value corresponding to a parametric gain $e^{2 r_{\rm max}} = 10.86$ dB (i.e.~$r_{\rm max} = 1.25$).  Shown is the Wigner function of the final cavity state; the structure corresponds to the expected entangled Rabi model ground state written in Eq.~(\ref{eq:ECS}).  (b) Final state after the same evolution period, but with no parametric drive and coupling enhancement; the Wigner function corresponds to the trivial weak coupling ground state.  (c)  Time-evolution of the logarithmic negativity $E_N(t)$, the fidelity $F(t)$ (c.f.~\cref{eq:F(t)}), and the parametric drive amplitude $\lambda(t)$ for a parametric gain of $10.86$ dB.  $E_N$ approaches its maximum value of $1$.  For all plots , $\gamma = 5\times 10^{-5} \delta_c$, $\kappa = 10^{-4} \delta_c$, and $t_f \simeq 5 \tau$. For an experimentally realistic detuning of $\delta_c = 1$ GHz, parameters correspond to $g = 100$ MHz, a cavity decay rate $\kappa = 100$ kHz, and a qubit lifetime $1/\gamma = 20\ \mu$s.}
    \label{fig:Entangle}
\end{figure}
%%%%%%%%%%%%%%%%

\cref{fig:Entangle} summarizes the the results of a simulation of our scheme for a system with $g=0.1 \delta_c$, where the system starts in the zero-coupling ground state $\ket{0} \ket{-z}$, and the protocol time scale is $\tau = 10  \delta_c^{-1}$.  Panel (b) shows the Wigner function of the cavity state obtained if the parametric drive is off during this evolution time:  it corresponds to vacuum.  Panel (a) shows instead the Wigner function obtained when the parametric drive is ramped such that $e^{2r_{\rm max}} = 11$ dB.   One clearly sees the double-blob structure associated with the target state in Eq.~(\ref{eq:ECS}), and \cref{fig:Entangle}(c) shows that one indeed has good fidelity with this state, with the expected near-maximal amount of qubit-cavity entanglement (as characterized by the log negativity $E_N$).

\begin{figure}[t]
    \centering
    \includegraphics[width=0.4\textwidth]{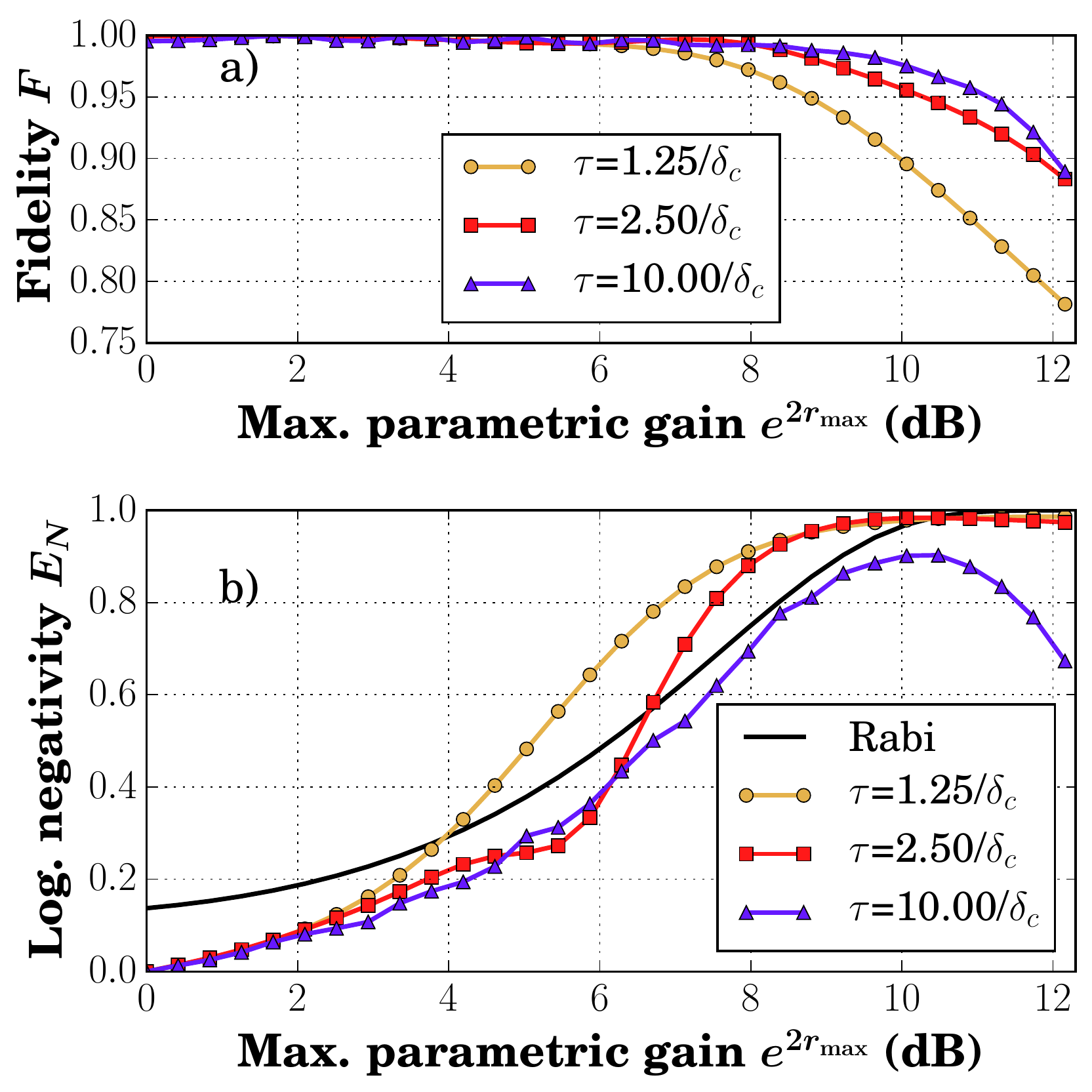}
    \caption{(a) Final state fidelity $F(t_f)$ (c.f.~\cref{eq:F(t)})  and (b) entanglement (measured via the logarithmic negativity $E_N$)
	as a function of the final parametric gain  $e^{2r_{\rm max}}$; different curves are for different values of the protocol speed parameter $\tau$. 		The total time of each simulation is $t_f \simeq 5 \tau$.
	The black curve in (b) is $E_N$ in the ground state of the ideal  Rabi-model Hamiltonian (c.f. $\hat{H}_{\rm Rabi}$ in \cref{eq:HSt}) with
	equivalent parameters: coupling $g e^{r_{\rm max}}/2$, cavity frequency $\delta_c \sech 2r_{\rm max}$ and qubit frequency $\delta_q$.
	Other simulation parameters are the same as in \cref{fig:Entangle}.}
    \label{fig:linecuts}
\end{figure}

It is also interesting to consider the performance of the adiabatic state preparation as a function of the parametric gain $e^{2 r_{\rm max}}$.
 \cref{fig:linecuts}(a) shows the behaviour of the fidelity $F(t_f)$ at the end of the protocol, as a function of the gain; different curves are for different ramp rates $1/\tau$.  The fidelity is seen to generically drop off at high amplification factors.  This is due to increased non-adiabatic errors (due to $\hat{H}_{\rm DA}$ in Eqs.~(\ref{eqs:Hams})), as well as due to dissipation.  In the lab frame, the cavity is driven by vacuum noise associated with the loss $\kappa$.  However, in the squeezed frame used to write Eq.~(\ref{eq:HSt}), this noise appears squeezed.  This unwanted squeezing is oriented such that it enhances the spurious terms in $\hat{H}_{\rm Err}$ in Eq.~(\ref{eq:HSt}), and also has effects akin to heating; this all leads to errors in the adiabatic protocol.
 \cref{fig:linecuts}(b) shows the corresponding behaviour of the qubit-cavity entanglement (measured by the log negativity $E_N$).  Surprisingly, the entanglement does not simply mirror the behaviour of the fidelity, and for rapid protocols, can even be larger than in the ideal target state.

%%%%%%%%%%%%%%%
% Fig. 5:  More entanglement details
%%%%%%%%%%%%%%%%
%%%%%%%%%%%%%%%%%%%

%%%%%%%%%%%%%%%%%%%%%%%%%%%%%%%%%%%%%%%%%
\emph{Conclusion}--We have analyzed how parametric driving of a cavity can enable a strong coupling enhancement in cQED, even letting a weak coupling system reach the regime of ultra-strong coupling.  The time-dependent control of the enhancement allows a variety of protocols, including the adiabatic preparation of strong-coupling highly entangled states.
Our scheme is well-suited for contemporary circuit QED technology, where strong parametric interactions \cite{Siddiqi2004,Yamamoto:2008aa}, and high coherence times \cite{Chen2014,Reagor2016} are commonplace. It can also be implemented in other cavity QED architectures, including ones incorporating nitrogen-vacancy centers \cite{Kubo:2010fk}, Rydberg atoms \cite{Pritchard:2014qf}, or quantum dots \cite{Mi:2017aa,Stockklauser2017}. Additionally, our scheme can be generalized to realize regimes of ultra-strong coupling in lattices of cQED cavities \cite{Houck2012}, by introducing local parametric driving at each site or a subset of sites, and can be applied to generate multi-qubit entanglement in a multi-qubit, single mode setup \cite{Armata:aa}.

\begin{acknowledgements}
This work was supported by NSERC and the AFOSR MURI FA9550-15-1-0029.
\end{acknowledgements}

\bibliography{references}

\pagebreak

\clearpage

\widetext
\begin{center}
\textbf{\large Supplemental Material for ``Enhancing cavity QED via anti-squeezing:  synthetic ultra-strong coupling''}
\end{center}
\setcounter{equation}{0}
\setcounter{figure}{0}
\setcounter{table}{0}
\setcounter{page}{1}
\makeatletter
\renewcommand{\theequation}{S\arabic{equation}}
\renewcommand{\thefigure}{S\arabic{figure}}
\renewcommand{\bibnumfmt}[1]{[S#1]}

\section{Time evolution}

In this section, we calculate the explicit form of $\alpha(t)$ used in Eq.~(5) of the main text. In the squeezed frame, the time-dependent coherent evolution of the system is described by the propagator
\begin{align}
	\hat{K}^{\rm S}(t) = \mathcal{T}\exp\left(-i\int_0^t \hat{H}^{\rm S}(t')dt'\right),
\end{align}
where $\mathcal{T}$ is the time-ordering operator. This can be expressed as
\begin{align}
	\hat{K}^{\rm S}(t) = \hat{K}_{\rm Rabi}(t) \mathcal{T}\exp\left(-i\int_0^t (\hat{H}_{\rm Err}^{\rm Rabi}(t')+\hat{H}_{\rm DA}^{\rm Rabi}(t'))dt'\right),
\end{align}
where
\begin{align}
	\hat{K}_{\rm Rabi}(t) = \mathcal{T}\exp\left(-i\int_0^t \hat{H}_{\rm Rabi}(t')dt'\right),
\end{align}
is the propagator for the target Rabi Hamiltonian evolution and
\begin{align}
	\hat{H}_{\rm Err/\rm DA}^{\rm Rabi}(t)=\hat{K}^{\dagger}_{\rm Rabi}(t)\left(\hat{H}_{\rm Err/DA}(t)\right)\hat{K}_{\rm Rabi}(t),
\end{align}
lead to small corrections to the ideal evolution. For $\delta_q = 0$ the Rabi propagator can be further simplified via a Magnus expansion,
\begin{equation}
\hat{K}_{\rm Rabi}(t) = \exp\left(\left(\alpha(t)\hat{a}^{\dagger}-\alpha^*(t)\hat{a}\right)\left(\hat{\sigma}_++\hat{\sigma}_-\right)\right) \exp\left(-i\Lambda_c(t,0)\hat{a}^{\dagger}\hat{a}\right), \label{eq:RabiPropa}
\end{equation}
where $\Lambda_c(t,t') = \int_{t'}^{t}\Omega_c(t'')dt''$ is the integrated cavity frequency and
\begin{equation}
\alpha(t) = \frac{g}{2i} \int_0^t \exp\left(r(t')-i\Lambda_c(t,t')\right) dt', \label{eq:alpha(t)}
\end{equation}
is the cavity displacement in the squeezed frame. In the adiabatic limit $\dot{r}(t) \approx 0$, \cref{eq:alpha(t)} reduces to the optimal displacement $ge^{r(t)}/2\Omega_c(t)$. \Cref{eq:RabiPropa} then propagates the state $\ket{0,-z}$ to the entangled cat-state
\begin{equation}
\ket{\Psi_{\rm Rabi}(t)} = \frac{\ket{+\alpha(t),+x}-\ket{-\alpha(t),-x}}{\sqrt{2}}.
\label{eq:RGS}
\end{equation}
Close to the adiabatic limit (i.e. $\alpha(t) \sim g e^{r(t)}/2\Omega_c(t)$), $\abs{\abs{\hat{H}_{\rm Err}^{\rm Rabi}(t)}}\sim  g e^{-r(t)}\abs{{\rm Im}\left[\alpha(t)\right]}/2 \ll \abs{\alpha (t)}$, and $\hat{H}_{\rm Err}^{\rm Rabi}(t)$ can be safely ignored for the entire protocol as long as ${\rm Re}[\alpha(t)] \gg {\rm Im}[\alpha(t)]$. $\abs{\abs{\hat{H}_{\rm DA}^{\rm Rabi}(t)}} \sim \abs{\dot{r}(t) \alpha(t)}$, and in general this is only constrained by the stability condition $\abs{\dot{r}(t)}\ll \Omega_c(t)$. This shows the importance of remaining adiabatic, $\dot{r}(t)\approx 0$, as $\hat{H}_{\rm DA}^{\rm Rabi}(t)$ contributes to the growth of ${\rm Im}[\alpha(t)]$, and therefore of $\hat{H}_{\rm Err}^{\rm Rabi}(t)$. In the adiabatic limit $\dot{r}(t)\approx 0$ and \cref{eq:RGS} is always valid.

\section{Master Equation Derivation}

In this section we outline the derivation of the master equation in the lab frame found in the main text. We start with the full Hamiltonian including the system-environment interaction
\begin{align}
\hat{H}_{\rm Lab} = \omega_c \hat{a}^{\dagger}\hat{a} + \frac{\omega_q\hat{\sigma}_z}{2} - \frac{\lambda}{2} ( \hat{a}^{\dagger 2}e^{-i\omega_pt}+\hat{a}^2e^{i\omega_pt}) + g(\hat{a}^{\dagger}\hat{\sigma}_-+\hat{\sigma}_+\hat{a}) + \hat{H}_{\rm E} + \hat{H}_{\rm SE},
\end{align}
where we consider bosonic environments for the cavity and the qubit, such that
\begin{align}
	&\hat{H}_{\rm E} = \sum_{\nu}\omega^{c}_{\nu}\hat{b}_{\nu}^\dagger\hat{b}_{\nu} +  \sum_{\nu}\omega^{q}_{\nu}\hat{c}_{\nu}^\dagger\hat{c}_{\nu}, \\
	&\hat{H}_{\rm SE} = \sum_{\nu}J_{\nu}^c\left(\hat{a} + \hat{a}^\dagger\right)\hat{X}_{\nu}^c + \sum_{\nu}J_{\nu}^q\hat{\sigma}_x\hat{X}_{\nu}^q = \sum_{\nu}J_{\nu}^c\left(\hat{a} + \hat{a}^\dagger\right)\left(\hat{b}_{\nu} + \hat{b}_{\nu}^\dagger\right) + \sum_{\nu}J_{\nu}^q\hat{\sigma}_x\left(\hat{c}_{\nu} + \hat{c}_{\nu}^\dagger\right),
\end{align}
where $J_{\nu}^{c/q}$ are the system-environment coupling strengths. We focus only on transversal coupling between the qubit and the environment, as this is most detrimental to our scheme.

Moving to a frame rotating at $\omega_p/2$ for both the system and the environments, the system-environment interaction Hamiltonian becomes time-dependent
\begin{align}
\hat{H}_{\rm SE}(t) &= \sum_{\nu}J_{\nu}^c\left(\hat{a}e^{-i\frac{\omega_p}{2}t} + \hat{a}^\dagger e^{i\frac{\omega_p}{2}t}\right)\left(\hat{b}_{\nu}e^{-i\frac{\omega_p}{2}t} + \hat{b}_{\nu}^\dagger e^{i\frac{\omega_p}{2}t}\right) \\ &+ \sum_{\nu}J_{\nu}^q\left(\hat{\sigma}_-e^{-i\frac{\omega_p}{2}t} + \hat{\sigma}_+e^{i\frac{\omega_p}{2}t}\right)\left(\hat{c}_{\nu}e^{-i\frac{\omega_p}{2}t} + \hat{c}_{\nu}^\dagger e^{i\frac{\omega_p}{2}t}\right). \label{eq:HEint}
\end{align}
However, as $\omega_p$ is much larger than any system frequency, we can make a rotating-wave approximation and drop counter-rotating terms to arrive at
\begin{align}
\hat{H}_{\rm SE}(t) = \sum_{\nu}J_{\nu}^c\left(\hat{a}\hat{b}_{\nu}^\dagger + \hat{a}^\dagger\hat{b}_{\nu}\right) + \sum_{\nu}J_{\nu}^q\left(\hat{\sigma}_-\hat{c}_{\nu}^\dagger + \hat{\sigma}_+\hat{c}_{\nu} \right). \label{eq:HEintRWA}
\end{align}
Crucially, in this frame the bath contains bosonic modes at negative frequencies down to $-\omega_p/2$
\begin{align}
	\hat{H}_{\rm E} = \sum_{\nu}\left(\omega^{c}_{\nu}-\frac{\omega_p}{2}\right)\hat{b}_{\nu}^\dagger\hat{b}_{\nu} +  \sum_{\nu}\left(\omega^{q}_{\nu}-\frac{\omega_p}{2}\right)\hat{c}_{\nu}^\dagger\hat{c}_{\nu}.
\end{align}
As a result, for this driven-dissipative system, the environment can mediate transitions from a lower energy system eigenstate to a higher energy eigenstate, even at zero temperature. This is in contrast to the undriven ultra-strong coupling situation, where such transitions are forbidden \cite{Beaudoin11}.

More formally, we consider diagonalizing the system self-Hamiltonian in the lab frame (rotating at $\omega_p/2$), such that
\begin{align}
	\hat{H} = \sum_{j}\epsilon_{j}\ketbra{j}{j},
\end{align}
where $\epsilon_j$ are the eigenfrequencies, and $\{\ket{j}\}_j$ the eigenstates of the system. Moving to the interaction picture with respect to the system and environment self-Hamiltonians the system opertors can be written in the interaction picture as
\begin{align}
	&\hat{a}^{\rm I}(t) = \sum_{j,k} \ketbra{j}{k}\bra{j}\hat{a}\ket{k}e^{i\epsilon_{jk}t}, \\
	&\hat{\sigma}_-^{\rm I}(t) = \sum_{j,k} \ketbra{j}{k}\bra{j}\hat{\sigma}_-\ket{k}e^{i\epsilon_{jk}t},
\end{align}
where $\epsilon_{jk} = \epsilon_j - \epsilon_k$ is the transition frequency of the eigenstate transition $\ketbra{j}{k}$. As the environment contains negative frequency modes down to $-\omega_p/2$, and $\abs{\omega_p} \gg \abs{\epsilon_{jk}}$, the environment can mediate all eigenstate transitions, regardless of whether $\epsilon_{jk} > 0$ or $\epsilon_{jk} <0$, as the environment spectral density will be nonzero at all $\epsilon_{jk}$. In the standard case, where the system is not driven, only the transitions with $\epsilon_{jk} > 0$ have nonzero spectral density and are possible.

In addition, we consider Markovian environments with white spectrums, such that the transition rates for all eigenstate transitions will be the same. Combining this with the fact that all transitions are allowed, we can assign a single jump operator for the cavity and a single jump operator for the qubit, i.e. $\hat{a}^{\rm I}(t)$ and $\hat{\sigma}_-^{\rm I}(t)$, with decay rates $\kappa$ and $\gamma$ respectively. In the lab frame, these simply correspond to the jump operators $\hat{a}$ and $\hat{\sigma}_-$. From here, a standard derivation of the Lindblad master equation will obtain Eq.~(6) of the main text.

%%%%%%%%%%%%%%%%%%%%%%%%%%%%%%%%%%%%%%%
\section{Qubit Absorption Spectrum}

The absorption spectrum $S[\omega]$ of the qubit is proportional to the rate at which the qubit would absorb power from a monochromatic qubit drive at frequency $\omega$.  As is standard, it is defined via the two-time correlation function
\begin{align}
	S(\omega) = \int_{-\infty}^{+\infty} \expectationvalue{\hat{\sigma}_-(t)\hat{\sigma}_+(t)}_{ss} e^{-i\omega t} dt,
\end{align}
where $\expectationvalue{\bullet}_{ss}$ is the expectation value with respect to the steady state of the system $\rho_{ss}$. We use the quantum regression theorem to compute the spectrum,
\begin{align*}
	S(\omega) = \int_{-\infty}^{+\infty} {\rm Tr} \left[\hat{\sigma}_-(t)[\hat{\sigma}_+ \hat{\rho}_{ss}]\right] e^{-i\omega t} dt  = \int_{-\infty}^{+\infty} {\rm Tr} \left[\hat{\sigma}_- [\hat{\sigma}_+ \hat{\rho}_{ss}](t)\right] e^{-i\omega t} dt,
\end{align*}
which shows that the spectrum can be computed using the single-time correlation $\expectationvalue{\hat{\sigma}_-(t)}$ with an initial pseudo-state $\hat{\sigma}_+\hat{\rho}_{ss}$. We calculate the absorption spectrum numerically, and plot $\abs{S(\omega)}$ in the main text Fig.~2.

The simulations for the spectrum shown in Fig.~2 assume that the qubit is driven by squeezed vacuum noise in the lab frame, with a squeeze parameter which matches the value of $r$ determined by the parametric drive.  As a result, in the squeezed frame, $\rho_{ss}$ corresponds to vacuum, and the obtained spectrum corresponds to the zero temperature absorption spectrum of a strong-coupling cQED system.  If no external squeezing was injected into the qubit, in the squeezed frame it would appear that the system is being driven by squeezed radiation.  As a result, the absorption spectrum would correspond to a strong-coupling cQED system driven by squeezed light.  Such spectra are characterized by peak asymmetries, as has been studied previously \cite{Cabrillo1996}.

%%%%%%%%%%%%%%%%%%%

\end{document}